\documentstyle[psfig]{aa} 
\begin{document} 
\thesaurus{ 
		09.13.2; % ISM: molecules
		11.09.4; % Galaxies: ISM
		13.18.1; % Radio continuum: galaxies		 
		13.18.3; % Radio continuum: ISM		 
		13.19.1; % Radio lines: galaxies		 
		13.19.3 % Radio lines: ISM		 
             } 
\title{CO lines in high redshift galaxies: 
perspective for future mm instruments} 
   
\author{F. Combes\inst{1}, R. Maoli\inst{1,2}, \and A. Omont\inst{2}
	} 
\offprints{F. Combes} 
\institute{
	Observatoire de Paris, DEMIRM, 61 Av. de l'Observatoire, 
	F-75014, Paris, France 
	\and 
	Institut d'Astrophysique de Paris-CNRS, 98 bis Bd. Arago, F-75014 Paris, France
	} 
\date{Received November 4, 1998; accepted February, 1999} 
\authorrunning{Combes et al..} 
\titlerunning{CO lines at high redshift}
\maketitle 
 
\begin{abstract}   
Nearly 10 high redshift ($z>2$) starburst galaxies have recently been 
detected in the CO lines, revealing the early presence in the universe of
objects with large amounts of already-enriched molecular gas. The latter
has sufficient density to be excited in the high-level rotational
CO lines, which yield more flux, making easier
high-redshift detections; however the effect is not as strong
 as for the sub-millimeter and far-infrared dust continuum
emission. With the help of simple galaxy models, based on these
first detections, we estimate the flux in all CO lines expected for
such starbursting objects at various redshifts. We discuss the detection
perspectives with the future millimeter instruments.
\keywords{ 
	ISM: molecules --
	Galaxies: ISM --
	Radio continuum: galaxies --		 
	Radio continuum: ISM	--	 
	Radio lines: galaxies	--	 
	Radio lines: ISM
	}  
\end{abstract} 
 %________________________________________________________________ 
  
\section{Introduction} 
 The detection of high-redhifted ($z>2$) millimeter CO lines in the
hyperluminous object IRAS 10214+4724 
($z=2.28$, Brown \& Vanden Bout 1992, Solomon et al. 1992a), has opened
a new way of research to tackle the star formation history of the Universe.
 Although the object turned out to be highly gravitationally amplified, it
revealed however that galaxies at this epoch could have large
amounts of molecular gas, excited by an important starburst,
and sufficiently metal-enriched to emit detectable CO emission lines.
 The latter bring fundamental information about the cold gas component 
in high-z objects and therefore about the physical conditions of
the formation of galaxies and the first generations of stars. 
At high enough redshifts, most of the galaxy mass could be molecular.
 The main problem to detect this molecular component could be its low 
metallicity, but theoretical calculations have shown that in a violent 
starburst, the metallicity could reach solar values very quickly
(Elbaz et al. 1992).

After the first discovery, many searches for other candidates took place,
but they were harder than expected, and only a few,
often gravitationally amplified,
objects have been detected: the lensed Cloverleaf quasar 
H 1413+117 at $z=2.558$ (Barvainis et al. 1994), 
the lensed radiogalaxy MG0414+0534 at $z=2.639$ (Barvainis et al. 1998),
the possibly magnified object
BR1202-0725 at $z=4.69$ (Ohta et al. 1996, Omont et al. 1996a),
the amplified submillimeter-selected hyperluminous galaxies SMM02399-0136
at $z=2.808$ (Frayer et al. 1998), and SMM 14011+0252 at 2.565
(Frayer et al. 1999), and the magnified BAL quasar APM08279+5255,
at $z=3.911$, where the gas temperature derived from the CO lines is 
$\sim$ 200K, maybe excited by the quasar (Downes et al. 1999).
Recently Scoville et al. (1997b) reported the detection of the first 
non-lensed object at $z=2.394$, the weak radio galaxy 53W002,
and Guilloteau et al. (1997) the radio-quiet quasar BRI 1335-0417, at $z=4.407$,
which has no direct indication of lensing.
If the non-amplification is confirmed, these objects
 would contain the largest molecular contents known
(8-10 10$^{10}$ M$_\odot$ with a standard CO/H$_2$ 
conversion ratio, and even more 
if the metallicity is low). 
The derived molecular masses are so high that H$_2$ would constitute
between 30 to 80\% of the total dynamical mass (according to the unknown
inclination), if the standard CO/H$_2$ conversion ratio was adopted.
The application of this conversion ratio is however doubtful, and it is
possible that the involved H$_2$ masses are 3-4 times lower (Solomon 
et al. 1997). 

Obviously, the search of CO lines in high-z objects is still a challenge
for present day instrumentation, but this will rapidly change with
the new millimeter instruments planned over the world
(the Green-Bank-100m of NRAO, the LMT-50m of UMass-INAOE, 
the LSA-MMA (Europe/USA) and the
LMSA (Japan) interferometers). It is therefore interesting to
predict with simple models the detection capabilities, as a function
of redshift, metallicity or physical conditions in the high-z objects.
In particular, it would be highly interesting to detect not only 
the few exceptional amplified monsters in the sky, but also the 
widely spread normal galaxy population of the young universe.
Our aim here is to determine to which redshift it will be possible,
and with which instrument. A previous study has already modelled
galaxies at very high redshift (up to $z=30$) and concluded
that CO lines could be even more easy to detect than the continuum
(Silk \& Spaans 1997). Our models do not agree with this conclusion.

Section 2 describes the two-component model we use for the molecular clouds
of the starburst galaxies, section 3 the cosmological model; both
are combined and the results are discussed in section 4.
Section 5 describes the detection perspectives with the future
millimeter instruments. 
%__________________________________________________________________ 
  
\section{Galaxy ISM modeling }  

 Since the physical conditions of the interstellar medium
is still unknown in early galaxies, the most straightforward
modeling is to extrapolate what we know from the local 
star-forming clouds in the Milky Way. Solomon et al. (1990)
showed that the ISM of ultra-luminous starbursting galaxies
is likely to contain a large fraction of dense gas,
corresponding to the star-forming cores of local molecular
clouds. The gross features of the CO and far-infrared emissions
of these luminous starbursts  can be explained by a 
dense (average density 10$^4$ cm$^{-3}$) molecular component
confined in the central kpc, containing cores of even
higher density ($\sim$ 10$^6$ cm$^{-3}$). These are about
100 times denser than in a normal galactic disk. The FIR to
CO luminosity ratio are nearly those expected for an optically
thick region radiating as a black-body (Solomon et al. 1997).
Typical masses of molecular gas are 2-6 10$^{10}$ M$_\odot$.

The dominant dust component in ultra-luminous galaxies has a 
temperature between 30 and 50 K, while there is sometimes
a higher temperature emission bump (e.g. Klaas et al. 1997).
The emission peaks around 100$\mu$m with a flux of 1-3 Jy
at $z = 0.1$. At the higher redshifts of $z =4-5$, continuum 
fluxes of a few mJy have been detected at 1.25mm (e.g. 
Omont et al. 1996b), corresponding to dust emission at 
about 220$\mu$m, with gas masses up to 10$^{11}$ M$_\odot$ if
no gravitational lens is present. As for the CO lines,
they are more difficult to detect with present day 
instrumentation, and most of the 8 detections reported so far
are gravitationally magnified. The flux detected
range between 3 to 20 mJy for lines (3-2) to (7-6)
(see Table \ref{COdata}).

\begin{table}[h]  
\caption[ ]{CO data for high redshift objects}
\begin{flushleft}  
\begin{tabular}{lllclcl}  \hline
Source    & $z$   &  CO  & S  & $\Delta$V& MH$_2$   & Ref  \\ 
          &       &line  & mJy  & km/s & 10$^{10}$ M$_\odot$    &           \\ 
\hline
F10214+4724 & 2.285 & 3-2  & 18 & 230  & 2$^*$      &  1   \\  
53W002      & 2.394 & 3-2  &  3 & 540  & 7          &  2   \\  
H 1413+117  & 2.558 & 3-2  & 23 & 330  & 2-6        &  3   \\  
SMM 14011+0252&2.565& 3-2  & 13 & 200  & 5$^*$      &  4   \\
MG 0414+0534& 2.639 & 3-2  &  4 & 580  & 5$^*$      &  5   \\  
SMM 02399-0136&2.808& 3-2  &  4 & 710  & 8$^*$      &  6   \\  
APM 08279+5255&3.911& 4-3  &  6 & 400  & 0.3$^*$    &  7   \\  
BR 1335-0414& 4.407 & 5-4  &  7 & 420  & 10         &  8   \\  
BR 1202-0725& 4.690 & 5-4  &  8 & 320  & 10         &  9   \\  
\hline    
\end{tabular} 
\end{flushleft}   
$^*$ corrected for magnification, when estimated\\
Masses have been rescaled to $H_0$ = 75km/s/Mpc. When multiple images
are resolved, the flux corresponds to their sum\\
(1) Solomon et al. (1992a), Downes et al (1995); (2) Scoville et al. (1997b); (3) Barvainis et 
al (1994, 1997); (4) Frayer et al. (1999);
(5) Barvainis et al. (1998); (6) Frayer et al. (1998); (7)
Downes et al. (1999); (8) Guilloteau et al. (1997); (9) Omont et al. (1996a)
\label{COdata}
\end{table} 

A striking feature of these ultra-luminous infra-red
objects, is that  both the IR and CO emissions originate in regions a few
hundred parsecs in radius (Scoville et al. 1997a, Solomon et al. 1997,
Barvainis et al. 1997).
Observations at arcsecond resolution have shown that the molecular
gas is confined in compact sub-kpc components, or disks of a few
hundred parsecs in radius. The average H$_2$ surface density there
is of the order of a few 10$^{24}$ cm$^{-2}$ (Bryant \& Scoville 1996). 
The gas must be clumpy, since molecules with high dipole moment are excited,
such as HCN (Solomon et al. 1992b). This strong central concentration
corresponds very well to what is expected in a merger by N-body
simulations (e.g. Barnes \& Hernquist 1992). Due to gas
dissipation and gravity torques,
large H$_2$ concentrations pile up at the galaxy nuclei in interacting
systems. Up to 50\% of the dynamical mass could be under
the form of molecular hydrogen in mergers  (Scoville et al.. 1991).
The most extreme 
ultra-luminous infrared galaxies, which are also mergers and
starbursts possess several 10$^{10}$ M$_\odot$ of H$_2$ gas, more than 
10 times the H$_2$ content of the Milky Way.

\subsection{A two-component model}

We will therefore base our simple model on a small inner
disk of 1 kpc diameter, containing two
density components (cf Table \ref{model}).
Both are distributed
in clouds of low volume filling factor. 
The dense and hot component represents the star-forming
cores, and for the sake of simplicity, there is one
core in every cloud. Because of
the large velocity gradients in the inner regions
(due to rotation or macroscopic velocity dispersion), 
most of these cores are not overlapping at a given velocity,
and their molecular emission can be simply summed up.
This is not the case for the more extended 
clouds (see below), or for the dust, that
could be optically thick at some frequencies. 
For the opacity of the dust, we take the formula computed
by Draine \& Lee (1984) and quite compatible with observations
(see Boulanger et al. 1996): 
$$
\tau = N_H(cm^{-2}) 10^{-25} (\lambda/ 250\mu)^2
$$

The total molecular mass is chosen to be
6 10$^{10}$ M$_\odot$, and the average column density N(H$_2$) 
of the order of 10$^{24}$ cm$^{-2}$ (typical of the Orion cloud
center). To fix ideas, we assume that the total mass
is distributed in 8.6 10$^7$ clouds of 700 M$_\odot$
each, with an individual velocity dispersion of 10 km/s,
embedded in a macroscopic profile of 300 km/s dispersion/rotation
(cf Table \ref{model}).
The column density of each cloud 
is respectively 3 10$^{22}$ cm$^{-2}$ for the extended
component and 3 10$^{23}$ cm$^{-2}$ for the core.
For the sake of simplicity, we consider cubic clouds
(the cube size is indicated in Table \ref{model}),
and their mass is computed taking into account the
helium mass (total mass
larger than the H$_2$ mass by a factor 1.33).

\begin{table}[h]  
\caption[ ]{Parameters of the two-component model}
\begin{flushleft}  
\begin{tabular}{lll}  \hline
Parameter	& Hot comp. & Warm comp.  \\ 
\hline
n(H$_2$) cm$^{-3}$	&   10$^6$	&    10$^4$\\
sizes (pc)       	&   0.1 	&    1     \\
$\Delta$V (km/s)       	&   10    	&    10     \\
$T_K$ ($z$ = 0.1)        &  90.0          &    30.0   \\
$T_K$ ($z$ = 1.0)        &  90.0          &    30.0   \\
$T_K$ ($z$ = 2.0)        &  90.0          &    30.0   \\
$T_K$ ($z$ = 3.0)        &  90.0          &    30.0   \\
$T_K$ ($z$ = 5.0)        &  90.0          &    30.1   \\
$T_K$ ($z$ = 10.0)        &  90.0          &    33.7   \\
$T_K$ ($z$ = 20.0)        &  91.0          &    57.5   \\
$T_K$ ($z$ = 30.0)        &  98.2          &    84.6   \\
N(CO) cm$^{-2}$          & 3. 10$^{19}$     & 3. 10$^{18}$ \\
N(H$_2$) cm$^{-2}$          & 3. 10$^{23}$     & 3. 10$^{22}$ \\
f$_s^*$          & 1.              &  100. \\
f$_v^*$          & 0.03              &  0.03 \\
mass fraction            & 0.1               &  0.9  \\
\hline    
\end{tabular} 
\end{flushleft}   
$^*$ f$_s$: surface filling factor, f$_v$ velocity filling factor \\
$T_K$ increases with $z$ keeping $T_{dust}^6 - T_{bg}^6$ constant (see text)
\label{model}
\end{table} 

As for the temperatures of the two components, we choose 
for local clouds $T_K$ = 30K for the extended component, and
90K for the star-forming cores. 
We do not take into account here a hotter component, possibly
heated by the AGN, as seen in F10214+4724 (Downes et al. 1995)
or the Cloverleaf (Barvainis et al. 1997).
These temperatures were
increased at high redshift, to take into account the
hotter cosmic background: indeed the dust is
heated above the background temperature by the
UV and visible light coming from the stars.
Assuming the same star-formation
rate for the clouds, and the same fraction of the star light
reprocessed by the optically thin dust, 
with an opacity varying as $\lambda^{-2}$,
the quantity $T_{dust}^6 - T_{bg}^6$ is fixed. This
means that the difference between the energy re-radiated
by the dust ($\propto T_{dust}^6$), and the energy received 
from the cosmic background by the dust ($\propto T_{bg}^6$)
is always the same. This is the energy flux coming from the stars.
We assume that the density is so high that the gas is heated
efficiently by the dust, and $T_{dust}$ = $T_K$ (which 
maximises the line flux). This condition might not
be satisfied, and our CO line fluxes from the gas
could be somewhat optimistic (however not by a large factor).
 The corresponding temperatures as a function 
of redshift are displayed in Table \ref{model}.
  We have also varied this condition, to take into account
other dust properties: in some dust infra-red spectra,
the opacity dependence with frequency has a lower
power than $\nu^2$ (sometimes $\nu^{1.5}$ or $\nu$);
also the dust could be completely optically thick,
so that the quantity
$T_{dust}^4 - T_{bg}^4$ is conserved.
This relation gives the maximum possible temperatures,
and the results are discussed below. With these two simple models,
the dust temperatures are bracketed.

\begin{figure}
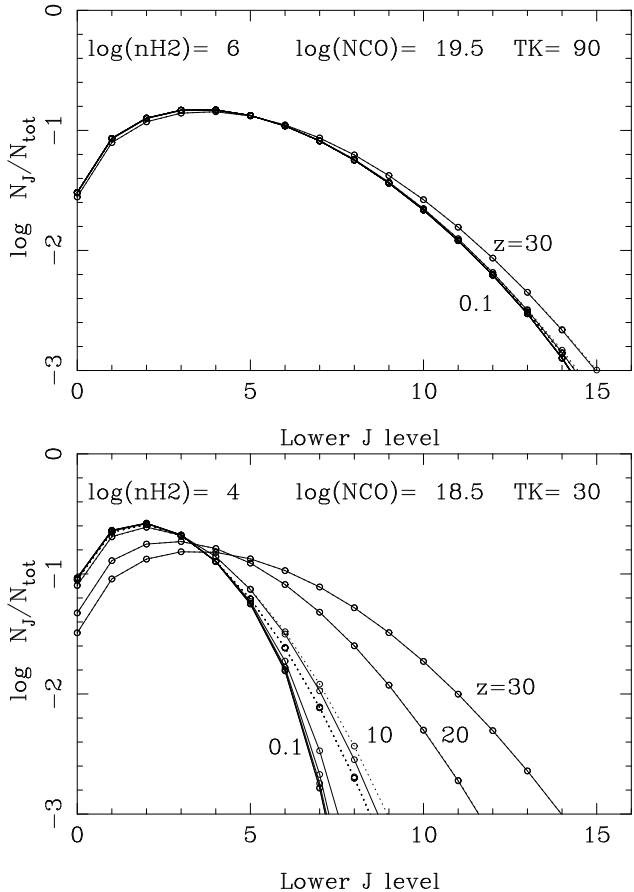

\psfig{width=8.5cm,file=8282_f1a.ps,bbllx=4cm,bblly=1cm,bburx=20cm,bbury=24cm,angle=-90}
\psfig{width=8.5cm,file=8282_f1b.ps,bbllx=4cm,bblly=1cm,bburx=20cm,bbury=24cm,angle=-90}
\caption{ Relative populations of the CO molecules, for the hot core (upper)
and warm cloud (bottom) components. The 8 curves correspond to redshifts
$z$= 0.1, 1, 2, 3, 5, 10, 20, 30. The values of
$T_K$ indicated are for $z$= 0 (see Table \ref{model}), and they do not vary
up to $z$ = 5. The full curves are the actual distributions, and
the dotted curves correspond to the LTE ones, at the corresponding $T_K$
(cf Table 2).
The 5 first dotted curves are superposed (same $T_K$), and the two last
ones ($z$ = 20 and 30) are coinciding with their corresponding full curves. }
\label{pop}
\end{figure}

We compute the relative populations of the CO molecule levels
in each component with the LVG approximation (using 17 levels); results are 
displayed in Fig. \ref{pop}, for each temperature  component.
At each redshift, the actual level populations are plotted as a full
line, and the LTE distribution as a dotted line for comparison.
 At high temperature, the full and dotted lines coincide.
From these distributions, excitation temperatures $T_{ex}$  and 
opacities $\tau$ can be derived for each CO line, and summed
with the relevant filling factors for the two components.
Antenna temperatures are obtained through
\begin{equation}
$$ T_A = [ f(T_{ex}) - f(T_{bg})] [1 - exp(-\tau)]$$
\end{equation}
where $f(T) = \frac{h\nu}{k}  (exp(\frac{h\nu}{kT}) -1)^{-1}$.
The flux is then 
\begin{equation}
$$ S = \frac{2 k T_A}{\lambda^2} \Omega_S$$
\end{equation}
where $\Omega_S$ is the angular size of the source.

The emission from individual clouds are summed directly, as long
as their total filling factor (i.e. the product of the surface
and velocity filling factor for the lines, or only the
surface filling factor for the continuum) is smaller than 1. 
When the total filling factor is larger than 1, 
the overlapping of clouds is taken into account, and
the opacity is increased on each line of sight by this factor.
This takes into account
the overcrowding and self-absorption of the cloud population.
 We computed the equivalent CO to H$_2$ conversion factor in this
model, for the CO(1-0) flux at $z = 0.1$: it is X= 5 10$^{20}$ 
mol/K/km.s$^{-1}$. At higher redshift, X increases, since 
the gas is hotter and the emission comes out in higher-$J$ lines than CO(1-0);
it is 9 10$^{20}$  at $z = 5$ and 280 10$^{20}$  at $z = 30$.

\subsection{An homogeneous sphere model}

 We also computed a quite different model of the molecular gas
in starburst, a homogeneous sphere model. This is motivated by the
consideration that the tidal forces and star formation energy
could be so high in the center as to disrupt molecular clouds.
The CO to H$_2$ conversion factor could then be quite different
from the standard one, based on a collection of virialised clouds.
 The total mass and size is the same in this model: 6 10$^{10}$
M$_\odot$ and 1 kpc diameter. The unique temperature is chosen to
be T$_d$ = 50K.  Since the surface density of the homogeneous source
is 3.5 10$^{24}$ cm$^{-2}$, the dust is optically thick for
wavelengths $\lambda  < $ 150$\mu$m, and the emission of the 
central molecular region can be approximately modelled
by a black-body (e.g. Solomon et al. 1997).
The average density is however 10$^{3}$ cm$^{-3}$, not enough to reach
LTE even in the low levels. The LVG model yields in this case an
excitation temperature approaching $T_K$ for the first CO levels,
then decreasing to a minimum, that could be 3 times below, and increasing
slightly for the highest levels (this translates into oscillations
of the flux versus wavelength, as shown in Fig. 7 below).
The equivalent CO to H$_2$ conversion factor in this
model, for the CO(1-0) flux at $z$ = 0.1, is now X= 3.7 10$^{20}$ 
mol/K/km.s$^{-1}$ ( 5.1 10$^{20}$  at $z$ = 5 and 370 10$^{20}$  at
$z$ = 30).

\begin{figure}
\psfig{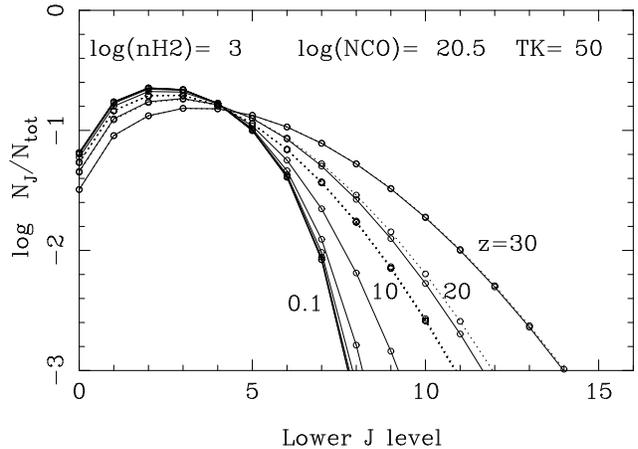}
\caption{ Same as Fig. \ref{pop}, for the homogeneous sphere model,
at $T_K$ = 50 K, at $z$ = 0. 
The 6 first dotted curves are superposed (same $T_K$), and the two last
ones ($z$ = 20 and 30) are almost coinciding with their corresponding full 
curves. }
\label{poph}
\end{figure}

% Optimum ?
It must be noted that the homogeneous gas model minimises the
optical depth of the CO lines, at a given velocity, since both
the spatial and velocity filling factors are 1, and consequently, it 
should maximise the amount of CO emission per gas mass. However, since
the corresponding density is then small (here 10$^{3}$ cm$^{-3}$),
the rotational levels are not excited to a $J$ as high as in the 
two-component clumpy model, and the excitation temperature is 
less than the kinetic temperature (cf Fig. \ref{poph}).
 This reduces the efficiency of the CO lines emission.
One way to have a better emissivity is to enlarge the 
starburst region to a size of several kpc, but this is not likely
to occur, given the observations of ultra-luminous galaxies
at low redshift, and the huge star formation rate then required.

In summary, the clumpy and homogeneous models are
two extreme simple models that help us to explore the large 
range of cloud distribution possibilities.

The principal way to optimize CO lines emissivity is to assume
that they are never highly optically thick, as assumed by 
Barvainis et al (1997). In their first interpretation of the CO lines
from the Cloverleaf, Barvainis et al (1994) derived an H$_2$ mass
of 4 10$^{11}$ M$_\odot$ (uncorrected for magnification), about an order
of magnitude larger than in Barvainis et al (1997). In the latter work,
the authors claim that the optical depth of the CO lines cannot be
much larger than 1, since the CO line ratios between the (3--2), (4--3),
(5--4) and (7--6) lines would then be Rayleigh-Jeans (in $\nu^2$),
which is not confirmed by observations. However, the various rotational
lines might not come from the same area, and their relative
magnification ratios are unknown. This prevents a definite conclusion
about the optical depth of the lines. Let us note that the low-optical
depth models are rather contrived, since to have a high enough excitation,
the density must be substantial, while the column density must be kept low.
 This results for instance in clumps of 0.1pc size for instance, in the
 T = 60 K model of Barvainis et al (1997), that completely cover
the source surface (with a low volume filling factor). To reach a diffuse
homogeneous sphere, the temperature should be larger than T = 300 K
(but then the expected dust emission is too large).

In our standard two-component clumpy model, the low-$J$ lines
are highly optically thick ($\tau \geq$ 100), and their relative intensities
are in the Rayleigh-Jeans ratio. We explore the optically thin cases in section
4, and in particular study the evolution of the continuum-to-line ratio.

\section{Cosmological model}  

\begin{figure}
\psfig{width=8.5cm,file=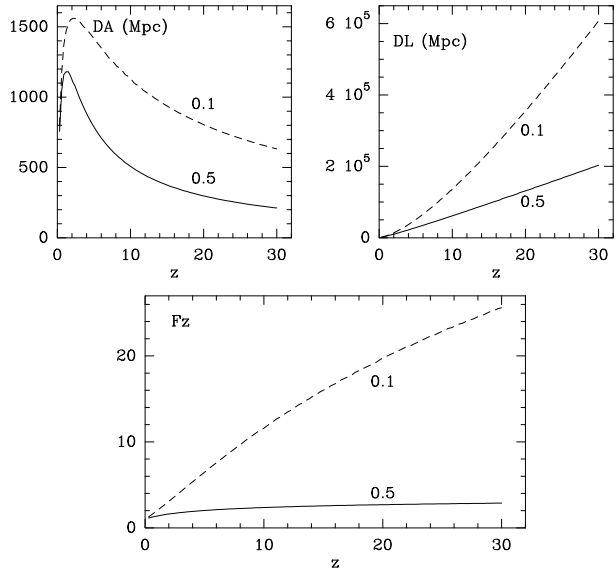,bbllx=15mm,bblly=5mm,bburx=20cm,bbury=21cm,angle=-90}
\caption{ Angular size distance ($D_A$), luminosity distance ($D_L$)
and correction factor $ F_z = (\frac{D_L H_0}{cz})^2$, for two values of
$q_0$ (0.5 full line, 0.1 dashed line).}
\label{distances}
\end{figure}

 For very distant ($z>1$) objects, the luminosity and angular size distances
are significantly different (cf examples in Fig. \ref{distances}).
  We parametrize as usual the cosmological model by the Hubble constant
$H_0$ (here fixed to 75 km/s/Mpc), and the deceleration parameter $q_0$.
We choose two values $q_0$= 0.1 and 0.5 to illustrate 
our computations here, corresponding respectively to an open universe, and 
to a critical (Einstein-deSitter model) flat one, for a zero  cosmological
constant (since in this case $\Omega_0$ = 2 $q_0$).
For a matter-dominated Friedmann universe,
the luminosity distance of an object at redshift $z$ is (e.g. Weinberg 1972):
\begin{equation}
$$ D_L = \frac{c}{H_0 q_0^2}[zq_0 + (q_0-1)(-1 + \sqrt{2q_0z+1})] $$
\end{equation}
and the angular size distance:
\begin{equation}
$$ D_A = (1 + z)^{-2} D_L $$
\end{equation}
As can be seen in Fig. \ref{distances}, the latter is decreasing with
$z$ as soon as $z>2$, i.e. the apparent diameter of objects increase with $z$.
 This may give the spurious impression that sources at high redshift will
be more easy to detect; in fact the measured integrated flux
decreases as  $D_L^2$.
 It is interesting to introduce the correction factor $F_z$, which is
the square of the ratio of the luminosity distance to the extrapolation of the 
low-redshift distance formula $cz/H_0$ (see for instance Gordon et al. 1992):
\begin{equation}
$$ F_z = (\frac{D_L H_0}{cz})^2$$
\end{equation}
This correction factor is also plotted in Fig. \ref{distances}, and 
is varying almost as $z$ for $q_0$ = 0.1. This means that for a
given intrinsic luminosity, this factor makes it more and more difficult
to detect objects at high redshift. The only favorable factor
in the millimetric domain is what is usually called 
``a negative K-correction''\footnote{The term of K correction may appear
confusing; it has been used for the first time by Wirtz in 1918,
for ``Kosmologie'' to determine the distance-redshift law; Hubble (1929)
called also K (after Wirtz's law) his now famous constant, and 
Oke \& Sandage (1968) quantified what they called the K-correction,
the combined effet of the changing $\lambda$ domain (if the spectrum of the 
object is not flat), and the reduced $\lambda$-interval observed at high
redshift, given a photometry band.}, i.e. that the flux
of the object increases with the frequency $\nu$ faster
than $\nu^2$ in the wavelength domain considered, and therefore
that its apparent luminosity could increase with redshift.  This is occuring 
for the dust emission from starbursts, which peaks 
in the 60-100$\mu$m region, and is progressively redshifted in the 
sub-millimeter and millimeter domain, where it is as easy to detect 
objects at $z=5$ than $z=1$ (Blain \& Longair 1993, 
McMahon et al. 1994, Omont et al. 1996b, Hughes et al. 1998).

To estimate detection capabilities, we will consider only point
sources, for the sake of simplicity.
At $z=2$, a beamwidth of 1$''$, reached already by present mm interferometers,
encompasses about 7 kpc, which is much larger than the area of dense 
and excited molecular gas in a starburst. With the foreseen next generation 
mm instruments, it will be possible to begin to resolve the emission
only for the best possible resolution (0.1$''$) and for very large redshifts, 
if however there is enough sensitivity.

At the present time, galaxies are detected in the optical up
to $z=$ 6 only. At this epoch, the age of the universe is about
5\% of its age, or 10$^{10}$ yrs in a standard flat model. 
For larger redshifts, it is likely that the total amount of
cumulated star formation is not a significant fraction of the total
(e.g. Madau et al 1996). However, it is of overwhelming interest
to trace the first star-forming structures, as early  as possible
to constrain theories of galaxy formation. At the recombination of
matter, at $z \sim$ 1000, the first structures to become unstable
have masses between 10$^5$ and 10$^8$ M$_\odot$, and at $z \sim$ 30,
it is possible than some structures of 10$^{10}$-10$^{12}$ M$_\odot$ become
in turn non linear, according to models, so we have computed our
estimations until such extreme redshifts ($\sim$ 30), even if such massive
objects are not likely to be numerous so early.

\section{Results and discussion}

\subsection{The standard clumpy model}

\begin{figure}
\psfig{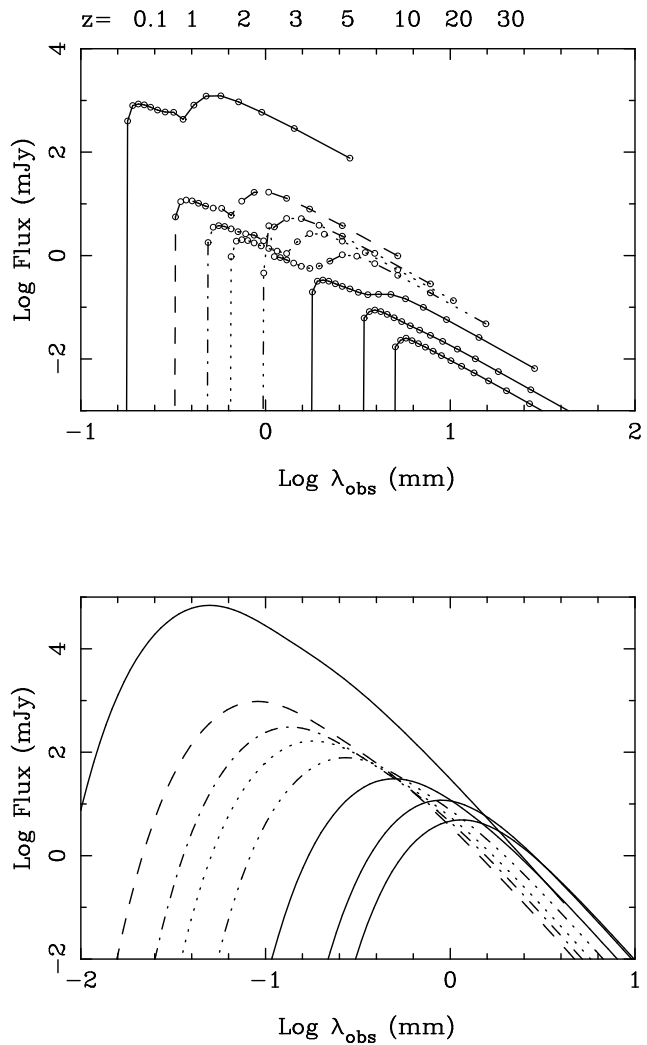}
\caption{ Expected flux for the two-component cloud model, for various
redshifts $z$ = 0.1, 1, 2, 3, 5, 10, 20, 30, and $q_0$ = 0.5.
. Top are the CO lines, materialised
each by a circle. Bottom is the continuum emission from dust.
It has been assumed here that $T_{dust}^6 - T_{bg}^6$ is conserved.}
\label{flux6}
\end{figure}

\begin{figure}
\psfig{width=8.5cm,file=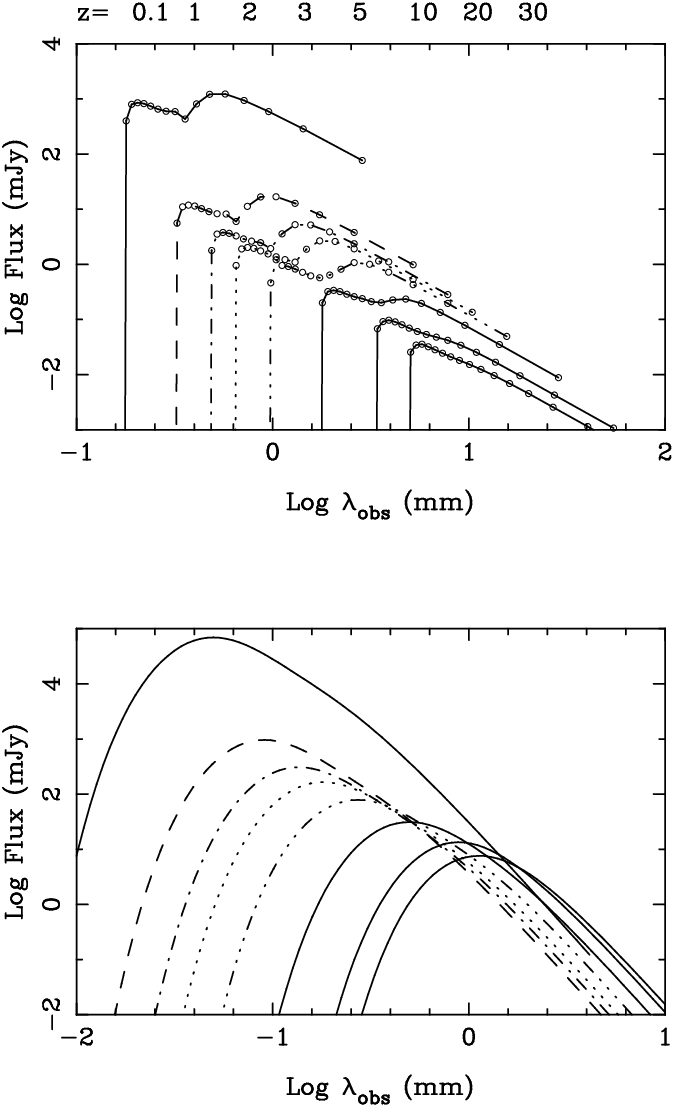,bbllx=15mm,bblly=5mm,bburx=21cm,bbury=12cm,angle=-90}
\caption{ Same as figure \ref{flux6}, but 
with $T_{dust}^4 - T_{bg}^4$ conserved, as for an optically
thick medium (see text for details).}
\label{flux4}
\end{figure}

The computed flux in the CO lines (not integrated in frequency) 
for the two-component cloud model is shown 
in Fig. \ref{flux6} for the 8 values of the redshift considered.
The two temperature contributions can
be seen clearly, although they tend to merge at high redshift.
The effect of the assumption on the gas temperatures can be seen
by comparison with Fig. \ref{flux4}, which maximises the gas
temperatures. We can see at once that
 the largest sensitivity for CO detection is 
around the lines CO(6-5) or CO(5-4) for the redshift up to $z$ = 5.
At larger redshifts, the higher lines CO(15-14) or CO(14-13)
will be the best choice, in the 3-5mm domain.

% commenter l'effet K-negatif
 The first effect to notice is the strong negative K-correction
in the dust emission,
which makes its detection even more easy at $z$ = 5 or 1. 
 For a fixed wavelength range, for instance at 3mm, we see that
the various redshift curves of Fig. \ref{flux6} are almost touching
each other, or are even $z$-reversed. However, the effect is
much more favorable for continuum detection. For CO lines,
we see more precisely that there is a factor 3 in flux between these
two redshifts ($z$ = 1 and 5) 
for a given $\lambda$, and therefore that ten times more
integration time is required to detect the same objects at $z$ = 5 
with respect to $z$ = 1. 
 This difference between the lines and continuum comes from the
fact that the lines are highly optically thick at long wavelengths:
the right side of the curves in Fig. \ref{flux6} for the lines 
go as $\lambda^{-2}$ (the Rayleigh Jeans approximation), while 
for the continuum, they go as $\lambda^{-4}$, due to the dust opacity
in $\nu^2$, and the fact that at low redshift and at these long wavelengths
the dust is optically thin. The K-correction advantage is then
much stronger for the dust emission; it is even optimum at very
high redshift ($z$ = 10 to 30), since the maximum dust emission then 
enters the mm domain, and begins to be optically thick.

\begin{figure}
\psfig{width=8.5cm,file=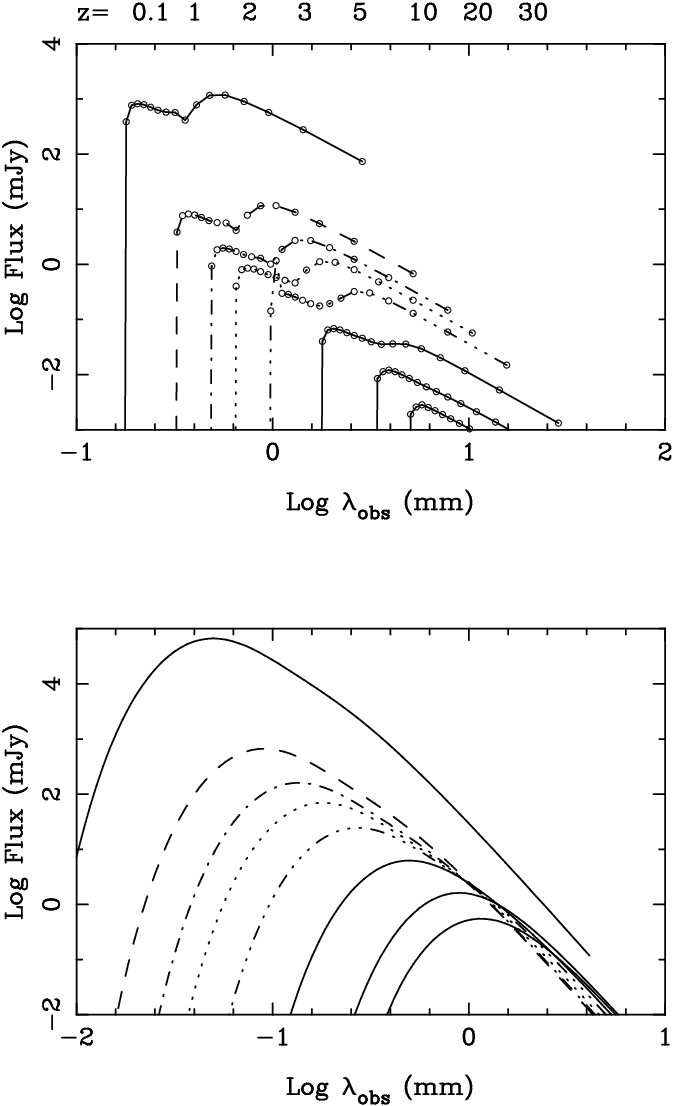,bbllx=15mm,bblly=5mm,bburx=21cm,bbury=12cm,angle=-90}
\caption{ Same as figure \ref{flux6}, but with $q_0$ = 0.1.}
\label{fluxq1}
\end{figure}

Fig. \ref{flux6} demonstrates that it is easier to detect the dust
emission of starbursts at $z > $10 than at $z$ = 5, at 1mm, and therefore
it should be possible to detect them with today instruments. However
all these estimates were computed for $q_0$ = 0.5. For $q_0$ = 0.1,
the fluxes are smaller at high redshift, as indicated in 
figure \ref{fluxq1}.

% commenter la difference de maxi entre raie et continuum
 A second obvious point to note, as far as the detection
of CO lines at high-$z$ is concerned, is that the maximum of
emission is always at longer wavelengths than for the continuum
(notice the different $\lambda$ scales in Fig. \ref{flux6}).
 In the case of the CO lines, the emission peaks at a frequency
lower by almost a factor 5 than in the case of the continuum.
 This is easy to understand, since it is the energy of the upper level $J$ 
of the transition which corresponds to the gas temperature; this energy
is proportional to $J (J+1)$. The energy of the transition is only
a fraction of it, proportional to $2 J$. Between the two, the ratio
is almost a factor 5, in the case of the CO molecule, excited at a
temperature of $\sim$ 90 K. The two peaks will be much closer
for molecules of higher rotational energy, such as H$_2$O for
instance, but their lines are expected to be much weaker and not
as favorable for detection (because of clumpiness and high optical depth).

% commenter sur l'excitation due au BG
A question that could be asked is whether the hotter cosmic background
at high $z$ helps in the detection of the CO lines.
 It can be seen on the relative populations of CO levels (Fig. \ref{pop})
that the density and column-densities we adopted for starburst objects
are always high enough to excite the levels nearly thermally (almost LTE).
 The effect of going to high redshift does not play on excitation
directly, but on the absolute temperature of the gas. The same effect 
occurs for the dust emission: temperatures are higher at high redshift,
for the same star formation rate. But this does not help {\it in fine}
on the detected flux level, since the observed flux takes into 
account the subtraction of the black-body emission itself (cf equation (1))
which is also higher at high $z$. The net effect for lines is even negative,
as shown in Fig. \ref{flux6}: the high-$z$ curves drop down from the
general tendency at lower redshift. This effect is generally
ignored when such estimations are done, and this is justified
for redshift $\leq$ 5 (e.g. Guiderdoni et al. 1998).  Very often,
for quick estimations, the curves of emission for a given object are simply
redshifted (translated in log-log plots) to estimate the K-correction
(e.g. Norman \& Braun 1996, Israel \& van der Werf 1996). 

\begin{figure}
\psfig{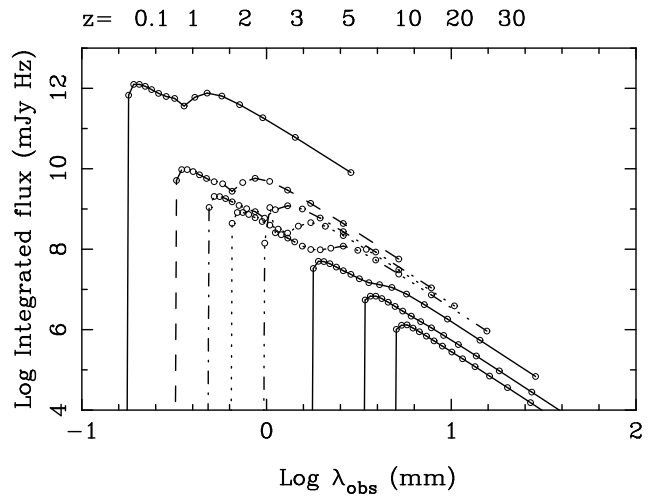}
\caption{ Same as figure \ref{flux6}, but for the integrated flux,
 $\int S_{\nu} d\nu$, more relevant for detection.}
\label{flux-int}
\end{figure}

% commenter integrated flux
One can also take into account that it is 
the {\it integrated} flux which is relevant to detect a line, since smoothing
proportionally to the line width increases the signal-to-noise.
Because the width of the line (at constant $\Delta V$) is proportional to
the frequency $\nu$, detection at high frequencies is then
favored, as shown in Fig. \ref{flux-int}.

\subsection{The homogeneous model}

\begin{figure}
\psfig{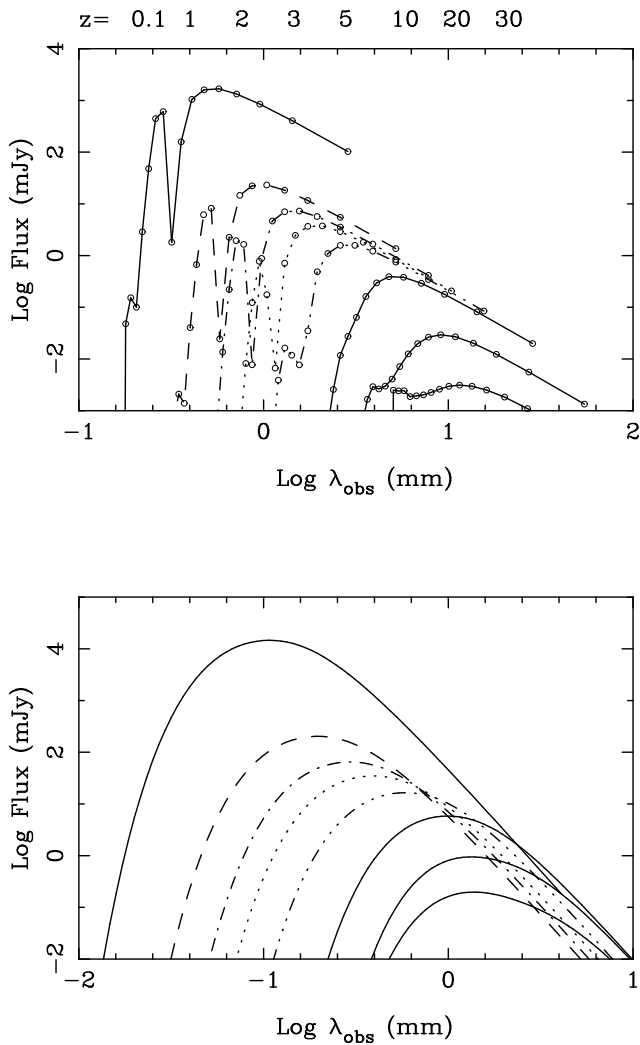}
\caption{  Same as figure \ref{flux6}, but with the homogeneous sphere
model, with T$_d$ = 50K. The non monotonous behaviour is due to excitation far
from LTE (see Section 2.2)}
\label{cohz_5}
\end{figure}

% commenter le modele homogene
In the case of the homogeneous sphere model, even though the CO to H$_2$
conversion factor is similar for the CO(1-0) line, our expectation
to detect such objects at high-$z$ is much less optimistic, 
because of the lower common 
temperature, and the lower excitation of the gas, which is now 
at low average density. The lines are significantly excited only
up to the CO(6-5), and this reduces the flux at high redshift,
for $z$ = 20 and 30 (see Fig. \ref{cohz_5}). The dust emission is also 
considerably reduced at high redshift.
Let us note that such low dust temperatures ($\leq 50K$) are relevant for a
fraction of high-$z$ sources such as BR1202-0725 (Cox et al. 1999), but
not for others such as F10214+4724 (Downes et al. 1995), the Cloverleaf
(Barvainis et al. 1997), or APM08279+5255 (Downes et al. 1999).

\subsection{Less optically thick models}

\begin{figure}
\psfig{width=8.5cm,file=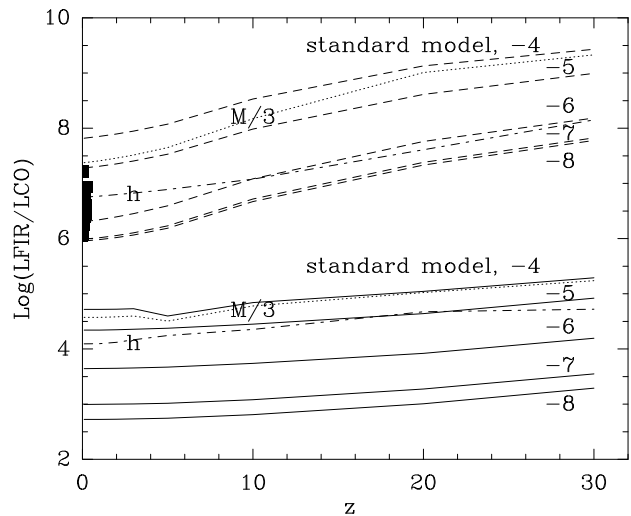,bbllx=5mm,bblly=3cm,bburx=18cm,bbury=18cm,angle=0}
\caption{ Luminosity ratio between the total Far infrared (FIR) dust emission
and the total integrated CO line emission (full lines), or only the integrated
CO(1-0) emission (dashed lines), as a function of redshift.
The lines are labeled by the log of CO/H$_2$ abundance ratio (-4 for
the standard model). The dotted lines indicate the result of dividing
the gas mass in the standard model by 3 (labeled M/3). The dash-dotted lines 
correspond to the homogeneous model (labeled h), displayed in fig 8. The ratio for the 37
ultra-luminous infrared galaxies observed by Solomon et al (1997) 
is marked as filled squares.}
\label{cont-line-ratio}
\end{figure}

 The standard two-component clumpy model is optically thick in most
CO lines. The optical depth comes essentially from the depth of
the cloud components themselves, but also from the
overlap of the clouds, at a given velocity. As can be seen in Table 
\ref{model}, the overlap amounts to $f_s f_v$ = 3 for the warm
component. It is therefore possible to have nearly the same CO emission
for about 3 times less gas mass, at least for the low-$J$ lines. 
The high-$J$ line emission is provided mainly by the core component,
for which there is no overlap ($f_s f_v$ = 0.03). There is no significant
absorption of the core emission by the warm component either,
since at the high frequencies of the main core contribution, 
the warm component is optically thin. Therefore, dividing the
total H$_2$ mass by 3, without modifying the cloud structure,
will result in about the same low-$J$ line emission, but 3-times
less high-$J$ emission. The continuum emission will also not be
simply divided by 3 (since the dust $\tau\ge$ 1 for $\lambda \le$ 200 $\mu$m),
but its spectrum will change shape. It is interesting to 
plot the continuum-to-line ratio for the various cases.
In figure \ref{cont-line-ratio}, we plot both the total dimensionless
L$_{FIR}$/L$_{CO}$ ratio, where L$_{CO}$ is integrated over all lines,
and the L$_{FIR}$/L$_{1-0}$ ratio, when only the CO(1-0) line is taken
into account, as is done observationally (e.g. Solomon et al 1997).
 We can see that both ratios increase with redshift, which confirms
the fact that the continuum will be easier to detect at high $z$
than the CO lines. The two cases, with 6 and 2 10$^{10}$ M$_\odot$
of gas have not very different L$_{FIR}$/L$_{CO}$ ratios
(although the L$_{FIR}$/L$_{1-0}$ ratio shows a marked effect, at
low redshift).

% on fait varier NCO par la metallicity
 Another obvious way to change the optical depth, without changing
the cloud structure, is to change the metallicity, and consequently
the CO and dust abundances. It is quite natural to keep the
cloud structures (and their high H$_2$ density) for starburst objects;
lower densities will not be able to excite the high-$J$ CO lines, 
penalizing detectability at high $z$. It is, however, likely that
gas at high $z$ is less enriched in metals and dust.
We have varied the CO/H$_2$ abundance ratio from 10$^{-4}$ in the
standard model to 10$^{-8}$ by factors of 10, keeping the total gas
mass at 6 10$^{10}$ M$_\odot$. Figure \ref{cont-line-ratio}
summarizes the results. For lower metallicities, 
since both CO lines and dust are then entirely
optically thin, the continuum-to-line ratio reaches a constant
value, independent of total mass. It is nearly two orders of magnitude
lower than in the standard model. 
 Note that our models bracket the range of observed values. The
standard model has a high ratio, due to it high gas mass, and high dust
temperature (the ratio varies at least as T$_d^3$, and even more for
moderate optical thickness).

\begin{figure}
\psfig{width=8.5cm,file=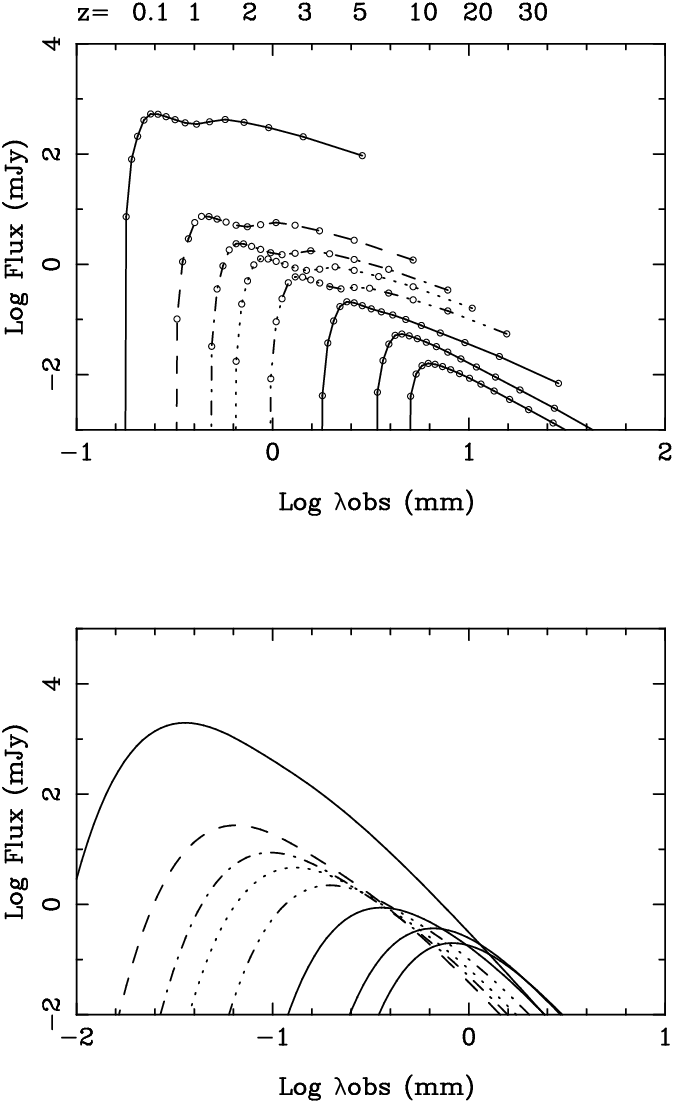,bbllx=15mm,bblly=5mm,bburx=21cm,bbury=12cm,angle=-90}
\caption{  Same as figure \ref{flux6}, but for one optically thin model,
with CO/H$_2$ = 10$^{-6}$.}
\label{thin}
\end{figure}

The resulting spectra for the optically thin model, with CO/H$_2$ = 10$^{-6}$,
are plotted in fig \ref{thin}. The same behaviour as a function
of redshift is observed, although the fluxes are lower, mainly 
in continuum.

\subsection{Conclusion}

In summary, we can see that the probability to detect objects at very high
redshift is much larger in the continuum dust emission than in the CO lines.
This is essentially due to a stronger K-correction advantage for the
continuum, due to the lower opacity of dust. However, the 
detection of CO lines brings a lot of 
complementary information, involving the redshift,
the line-widths and the kinematics of the gas. Also the line
detection suppresses the problems of confusion of sources, that
can affect the continuum surveys with single dishes.

We remark that we do not retrieve the surprising result
of Silk \& Spaans (1997), that the CO lines are  even easier
to detect than the continuum at very high redshift.
From our results, the continuum-to-line flux ratio increases with redshift
whatever the model and the optical thickness.
In figure 2 from Silk \& Spaans (1997), the maximum line
flux for the CO lines does not vary significantly
with redshift from 5 to 30, while we have in all cases
a variation  as large as (1+z)$^{-2.5}$ in the same range.\footnote{ 
We can note that they do not take into account the overlapping of
clouds on the line of sight, but this is in our model a factor 3 only,
and cannot account for the factor $\sim$ 60 discrepancy.}

\section{Detection perspectives}

We have computed, for both line and continuum, the integration time
required to detect a high-redshift object, with the best possible CO line,
or the best continuum frequency. These are displayed in Tables 
\ref{line-h} and \ref{cont-h}. A signal-to-noise ratio of a factor 5
has been assumed on the continuum flux $S_{\nu}$, and 
these estimations have been done
for the two-component model of Fig. \ref{flux6}.
For the line, the signal-to-noise ratio adopted is 3 when the
noise is smoothed over 30 km/s width, or 9 when smoothed over
the whole line width (300 km/s).
Since the sub-mm and mm domain is in fact punctuated by transparent
atmospheric windows and opaque broad regions, due to O$_2$ and H$_2$O,
some particular redshifts will be severely disfavored, as for the
CO lines detection. We have not taken into account such a complex
frequency dependence in our estimations, but on the contrary, we have 
kept only the upper envelope of atmospheric transmission
all over the domain (in table \ref{line-h}, we have however avoided
to estimate integration times exactly at the center 
of atmospheric lines). The final estimations are therefore valid
statistically for a big sample of sources, but can be wrong for
a given object for which the best CO lines fall in an opaque
atmospheric region. This is not as severe for continuum detections.

\begin{figure}
\psfig{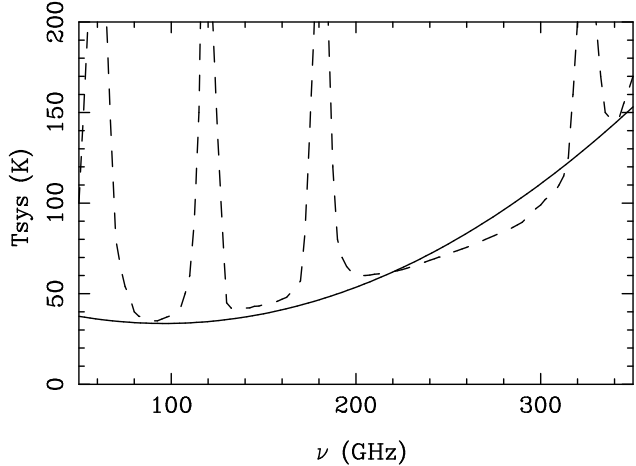}
\caption{ System temperature as a function of frequency adopted
for the LSA/MMA instrument. 
Dash curve: actual expected T$_{sys}$, corresponding to a receiver 
temperature equal to 2 h$\nu$/k, operating in full SSB mode,
with 1mm height of water, temperature 0$^{\circ}$ (at altitude 5 km);
Full curve: second-order polynomial fit}
\label{atm}
\end{figure}

For presently operating telescopes, the system temperature is 
taken from well known measurements between 1 and 3mm, and then
extrapolated as a second order polynomial in frequency.
For planned instruments, the large foreseen improvement in
receivers is taken into account (about a factor 4 in
system temperatures, and therefore noise power,
cf Guilloteau, 1996). 
This corresponds to receiver temperatures of the order of 2 $h\nu/k$, 
and SSB mode (at least 20 dB rejection). 
For the LSA-MMA project, a configuration of 64 antennae of 12m has been
adopted (providing a collecting surface of 7200 m$^2$). 
The expected surface efficiency of telescopes and their
altitudes are also taken into account.
Fig. \ref{atm} shows how the 2nd-order polynomial
fit corresponds to the lower envelope of the expected T$_{sys}$ for 
the LSA/MMA. For the continuum, the flux is smoothed
over 0.5 and 4 GHz for the IRAM and LSA/MMA interferometers
respectively, and 0.1 $\nu$ for bolometers. The sensitivity
figures given in Table \ref{cont-h} for the continuum receivers 
have been taken from Holland et al (1998) for JCMT-Scuba, 
and from Glenn et al (1998) for Bolocam.
All estimates have been done assuming point sources.

\begin{table*}[h]  
\caption[ ]{Smallest integration time to detect CO lines, (with optimum
$\nu$ in GHz)}
\begin{flushleft}  
\begin{tabular}{lll @{\hspace{1cm}}|| @{\hspace{1cm}} lll}  
\multicolumn{3}{c @{\hspace{1cm}}|| @{\hspace{1cm}} }{Present Receivers}
& \multicolumn{3}{c}{Future Receivers} \\
%   & Present   &Receivers &             & Future        & Receivers \\
   &&&&&\\
\hline 
   &&&&&\\
z  & IRAM-30m   & IRAM-PdB   & GBT-100m   & LMT-50m    & LSA-MMA  \\
   &&&&&\\
\hline 
   &&&&&\\
0.1& 3mn (209)  & 1mn (209)  & 0.4s (105) & 1.5s (209) & 0.04s (209) \\
1  &  36h (230) & 16h (230)  &  4mn (115) & 15mn (230) & 22s (230) \\
2  & 180h (230) &  70h (230) &  8mn (115) &1h15 (230)  & 2mn (230) \\
3  & 210h (144) &  86h (144) & 13mn (115) &1h15 (144)  & 2mn (144) \\
5  &  36d (115) & 14d (115)  & 1h (115)   & 10h (115)  & 15mn (115) \\
10  & --        & --         & 13h (115)  &  5d (146)  & 3h (146) \\
20  & --        & --         &  90h (77)  & --         & 28h (77) \\
30  & --        & --         &  54d (52)  & --         & 15d (52?) \\
\hline    
\end{tabular} 
\end{flushleft}   
The time is estimated to have a 3$\sigma$ detection of the flux
$S_{\nu}$, smoothed to 30 km/s \\
(one tenth of the profile width of 300 km/s), equivalent to a 9$\sigma$ 
detection, smoothed to 300 km/s \\
For all future receivers, $T_{sys}$ has been computed assuming
$T_{rec}$ = 2 h $\nu$/k, and 1mm height of water
\label{line-h}
\end{table*}

\begin{table*}[h]  
\caption[ ]{Smallest integration time to detect dust continuum 
(at 5$\sigma$, with optimum $\lambda$ in $\mu$m)}
\begin{flushleft}  
\begin{tabular}{lll @{\hspace{1cm}}|| @{\hspace{1cm}} lll}  
\multicolumn{3}{c @{\hspace{1cm}}|| @{\hspace{1cm}} }{Present Receivers}
&\multicolumn{3}{c}{Future Receivers} \\
%   & Present   &Receivers &             & Future        & Receivers \\
   &&&&&\\
\hline 
   &&&&&\\
z & IRAM-30m   & JCMT-Scuba &  CSO-Bolocam & LMT-Bolocam & LSA-MMA  \\
   &&&&&\\
\hline 
   &&&&&\\
0.1& 6mn (1250)&25s (350)  & 50s  (1100) &   0.1s (1100) &   36ms (850) \\
1  &6.7h (1250)&  1h (450) & 55mn (1100) &     6s (1100) &   2.7s (850) \\
2  &3.9h (1250)&52mn (850) & 35mn (1100) &     4s (1100) &   1.8s  (850) \\
3  &2.2h (1250)&38mn (850) & 20mn (1100) &   2.3s (1100) &   1.5s (850) \\
5  &1.1h (1250)&22mn (850) & 10mn (1100) &   1.2s (1100) &   0.9s (1250) \\
10 &24mn (1250)& 9mn (850) &  4mn (1100) &   0.5s (1100) &   0.3s (1250) \\
20 &13mn (1250)&19mn (850) &  3mn (1100) &   0.4s (1100) &   0.15s (1250) \\
30 &44mn (1250)&1.2h (1350)& 16mn (1100) &   1.8s (1100) &   0.6s (1250) \\
\hline    
\end{tabular} 
\end{flushleft}   
The following sensitivities were used, for one second integration time:
50mJy (IRAM-30m at 1250$\mu$m); \\
1200, 530, 80 and 60mJy (JCMT-Scuba, at 350, 450, 850 and 1350$\mu$m 
respectively); \\
30mJy (CSO-Bolocam at 1100$\mu$m); 1.3mJy (LMT-Bolocam at 1100$\mu$m); \\
2 and 0.7 mJy (LSA-MMA at 850 and 1250$\mu$m, respectively)
\label{cont-h}
\end{table*}

% Conclusions
The results in Tables \ref{line-h} and \ref{cont-h} are of course only
orders of magnitude, since they depend on the starburst model, and also
the technical performances of planned instruments are only extrapolated.
But they give already a good insight in what will be feasible in the
next decade. The continuum sources are already detectable with present
instruments at all redshifts. Large interferometers will be
necessary to reduce the confusion level, and map the sources.
The CO lines are presently not detectable easily at redshifts larger
than 1. The few detections already published at high $z$ owe their
detection to the high magnification factor provided by a gravitational lens,
or to exceptionally massive objects. For instance, the actual integrated
intensity of BR 1335-0414, 2.8 $\pm$ 0.3 Jy km/s at $z=$ 4.4,
Guilloteau et al 1997) is a factor 9 larger than our standard model
at $z=$ 5, at 3mm.
With the future instruments, they will be detectable easily up
to $z = 10$ and may be larger, if huge starbursts exist there.
At $z < 5$, it will be possible to detect CO lines from more normal
galaxies, and to tackle star formation in those very young galaxies.
 More exotic lines, such as CS, HCN, or even H$_2$O will then be available
to explore the physics of the interstellar medium 
and star formation in more detail.

\begin{acknowledgements}
We thank D. V. Trung for his LVG code, 
James Lequeux and F. Viallefond for useful discussions,
and an anonymous referee for helpful and detailed comments.
\end{acknowledgements} 
%  
%_____________________________________________________________________ 

\end{document}